\documentclass[12pt]{article}
\usepackage{t1enc}
\usepackage[latin1]{inputenc}
\usepackage{graphicx}
\usepackage[dvips]{rotating}
\usepackage{flafter}
\begin{document}

\begin{center}{\bfseries \large The \emph{Dauer Mutation} of the \emph{Caenorhabditis Elegans}, simulated with the Penna and the Stauffer Model}
\end{center}

\begin{center}{KERSTIN COLONIUS
\\Institute for Theoretical Physics, Cologne University, D-50923 K\"{o}ln, Germany}
\end{center}

\begin{abstract}
  Two ageing models were analysed whether they can confirm that the \emph{dauer mutation} of the \emph{nematode} helps to preserve the species. As a result the Penna model shows that populations with \emph{dauer larvae} survive bad environmental conditions, whereas populations without it die out. In the Stauffer model the advantage of the \emph{dauer mutation} for the survival is only given under certain conditions.
\end{abstract}
 
\begin{normalsize}


\end{normalsize}

\section{Introduction}
Some animals like the small nematode \emph{Caenorhabditis Elegans} have a mutation which is called \emph{dauer}. Their study may help our understanding of ageing in general. Finch and Kirkwood discovered that ageing worms get randomly damaged cells, like humans, \cite{Finch} and Herndon et al. claimed that old worms, like old people, suffer muscle decline \cite{Herndon}.
\\At the existence of a pheromone (a measure of population density), high temperatures or a food shortage at the end of the second larvae stage, the \emph{nematode} can go into a mutated third larvae stage being able to move but needing no food \cite{Wood}. This stage can be passed through 6 to 8 times until the conditions have improved \cite{Wood}. As the \emph{after-dauer} life span is not influenced by the endurance of the larvae stage, scientists agree that the \emph{nematodes} do not age during the \emph{dauer state} \cite{Wood}. This paper deals with the question if the \emph{dauer mutation} helps a population to preserve its species in bad times. For the Dasgupta model, this was already answered positively by Heumann \cite{Heumann}.  

\section{The \emph{dauer mutation} simulated with the asexual Penna model}
The 1995 developed asexual Penna model is a \emph{bit-string model}, in which each genome of an individual is represented by a computer word of 32 bits. A bit set to zero represents a healthy gene, whereas a bit set to one symbolises a hereditary disease that becomes active in the "year" (or other suitable time unit) represented by the position of the bit. Exceeds the number of active diseases a certain threshold, the individual dies \cite{Moss}. For further information about the asexual Penna model see e.g. \cite{Moss} or \cite{Penna}.
\\The original Penna model is now modified so that it contains an environmental condition after reaching a stable population. In this environmental condition each tenth summer is a bad one in which only 1\% of the population survive. After this bad tenth summer, as an option in the program, up to three more bad summers can follow one after the other, each with a probability of 0.5. Furthermore the \emph{dauer mutation} is included in the computer program so that all \emph{nematodes} of age 3 (representing the individuals in the third larvae stage) do not die in these bad summers. The other parameters were chosen as in the program listed in \cite{Moss}, if not stated otherwise. A summer means approximately one day in the \emph{Caenorhabditis Elegans'} life, as its mean life span is about 20 days \cite{Nuno}. In addition the \emph{dauer larvae} do not age in this simulation. 

\begin{figure}[h]
\centering
\includegraphics[width=9cm, angle=-90]{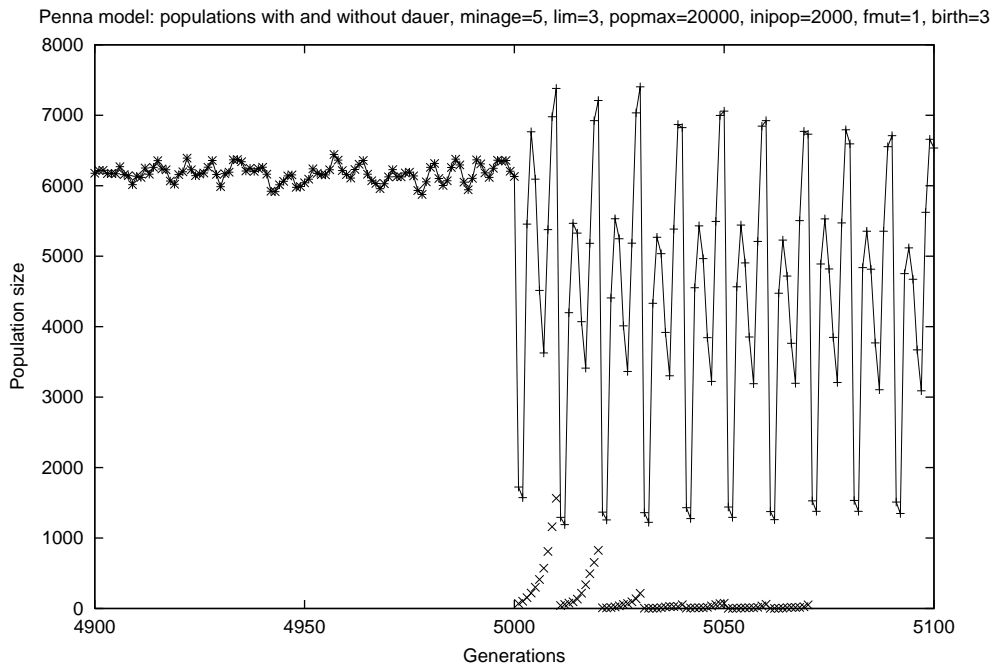}
\caption{After 5000 generations, in each tenth summer only 1 \% survive.
'+'=population with dauer, 'x'=population without dauer}
\end{figure}

\begin{figure}[h]
\centering
\includegraphics[width=9cm, angle=-90]{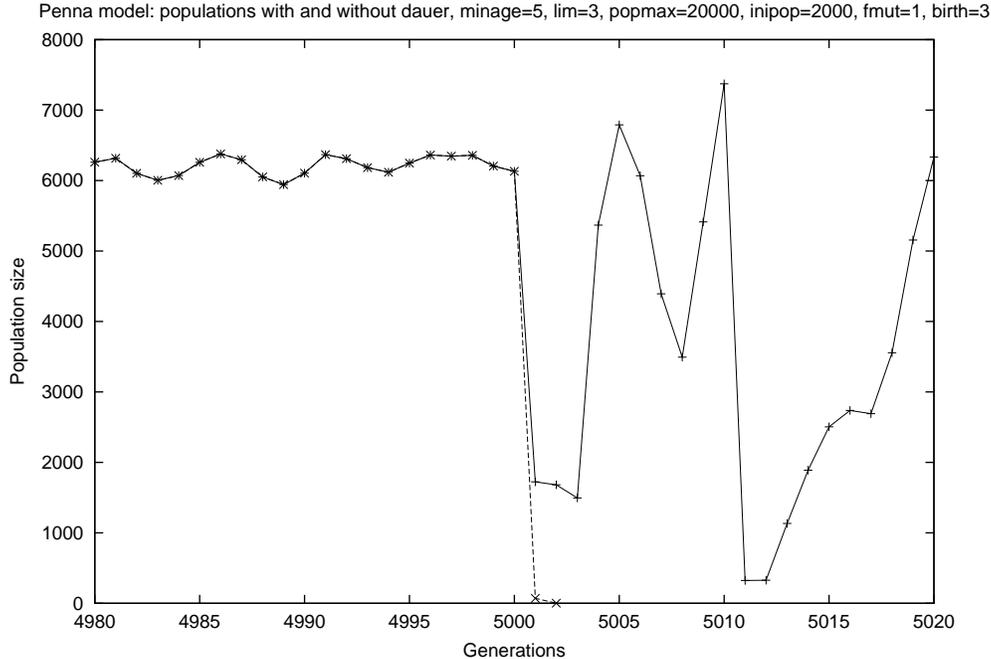}
\caption{After 5000 generations, each bad tenth summer may be followed by up to three bad ones.
'+'=population with dauer, 'x'=population without dauer}
\end{figure}

As a result figure 1 shows the comparison between a population with \emph{dauer} and one without it. The population size is plotted versus the generations. 
\\ The population with the \emph{dauer mutation} survives the bad conditions showing extreme fluctuations, whereas the population without the \emph{dauer larvae} dies out in the seventh bad summer.

Figure 2 shows the comparison between a population with and one without \emph{dauer larvae} at extreme conditions with up to four bad summers one after the other. This close-up shows the time between 4980 and 5020 generations. Again the population with \emph{dauer} survives, the population without it is already extinct in the second bad summer.

\section{The \emph{dauer mutation} simulated with the Stauffer model}
Stauffer suggests a simple alternative to the Penna model (\cite{Stauffer},\cite{Laessig}) that does not consider an explicit bit-string as genome. In this Stauffer model only the minimal reproduction age $a_{m}(i)$ and the genetic death age $a_{d}(i)$ are transmitted from generation to generation \cite{Meyer}. Having achieved the minimal reproduction age, the individual produces offspring with the probability \emph{b}
\[b=\frac{1+\epsilon}{a_{d}(i)-a_{m}(i)+\epsilon}\]
with the parameter $\epsilon=0.08$ to avoid divergences and extinction of the population \cite{Laessig}. Thus the birth rate is the smaller the longer the reproduction phase of the parent is: \emph{fecundity-survival trade-off} \cite{Stauffer}. The offspring inherits  $a_{m}(i)$ and  $a_{d}(i)$ from the parent with a mutation of $\pm 1$ ``year'' \cite{Stauffer}.
\\ The individuals reach at most their genetic death age \cite{Laessig}. They can die with a \emph{Verhulst probability } representing food and space restrictions as in the Penna model. The \emph{Verhulst survival probability} is given by $V=1-N/N_{max}$ with $N_{max}$ the capacity of the ecosystem \cite{Stauffer}.
\\Programming this model showed that a small difference in the interpretation leads to relevant effects on the results obtained. Whereas in Stauffers program the individual can get its first child with the age $> a_{m}$, in my program the reproduction starts already with reaching $a_{m}$. This difference results in a population that is about twice as high as in Stauffers version. The mortality in both versions consequently shows differences indicating a slightly better exponential increase in Stauffers interpretation. However both versions are regarded showing also differences in the simulation of the \emph{dauer larvae}. 
\\ In analogy to our simulation with the Penna model an environmental condition and the \emph{dauer state} are included. 
The parameter $s_b$ indicates how much of the population survives in the bad summer because of the environmental condition. In the case of the modified Penna model a disastrous summer was simulated: $s_b=0.01$. 
\\ Figure 3 shows populations with and without \emph{dauer mutation} for disastrous summers with $s_b=0.01$ and summers with $s_b=0.61$. It can be seen that all populations die out in Stauffers version, whether they have \emph{dauer larvae} or not, even when the summer with $s_b=0.61$ is not catastrophic. The \emph{dauer state} helps here very little.

\begin{figure}[h]
\centering
\includegraphics[width=9cm, angle=-90]{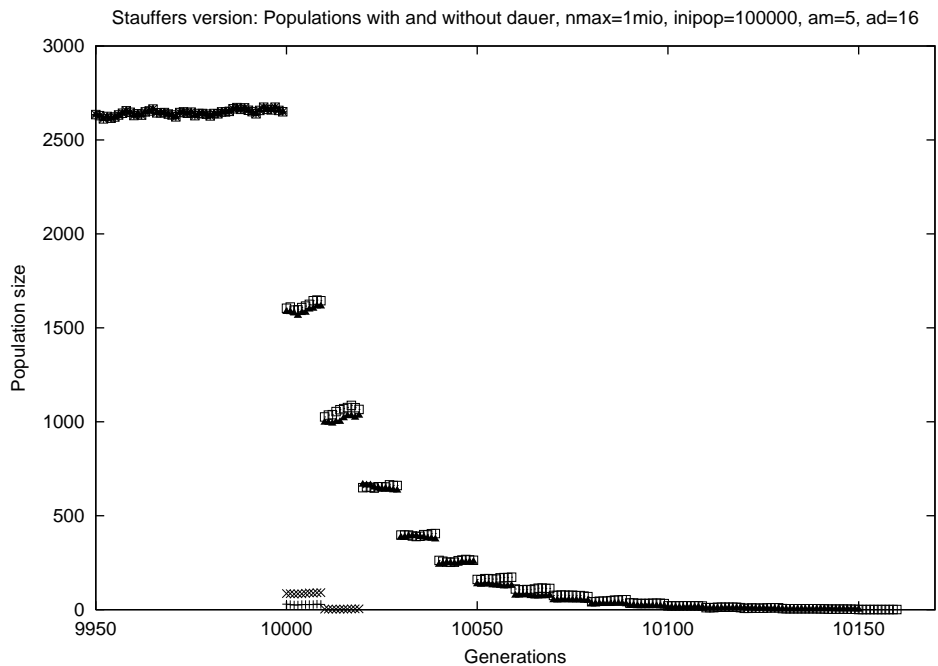}
\caption{After 10000 generations, each tenth summer is a bad one.
'x'=population with dauer,$s_b=0.01$, '+'=population without dauer, $s_b=0.01$,'square'=population with dauer, $s_b=0.61$, 'triangle'=population without dauer, $s_b=0.61$}
\end{figure}

Figures 4, 5 and 6 show my version with the same parameters as in figure 3. In my version both populations also die out for $s_b=0.01$, shown in figure 4. Figure 5 illustrates a population without \emph{dauer} dying out after a few bad summers at $s_b=0.61$, whereas the population with the \emph{dauer state} in figure 6 survives those conditions.
\\The comparison of the two versions shows that a small difference in the interpretation of the model can lead to totally different results concerning the preservation of the species.

\begin{figure}[h]
\centering
\includegraphics[width=9cm, angle=-90]{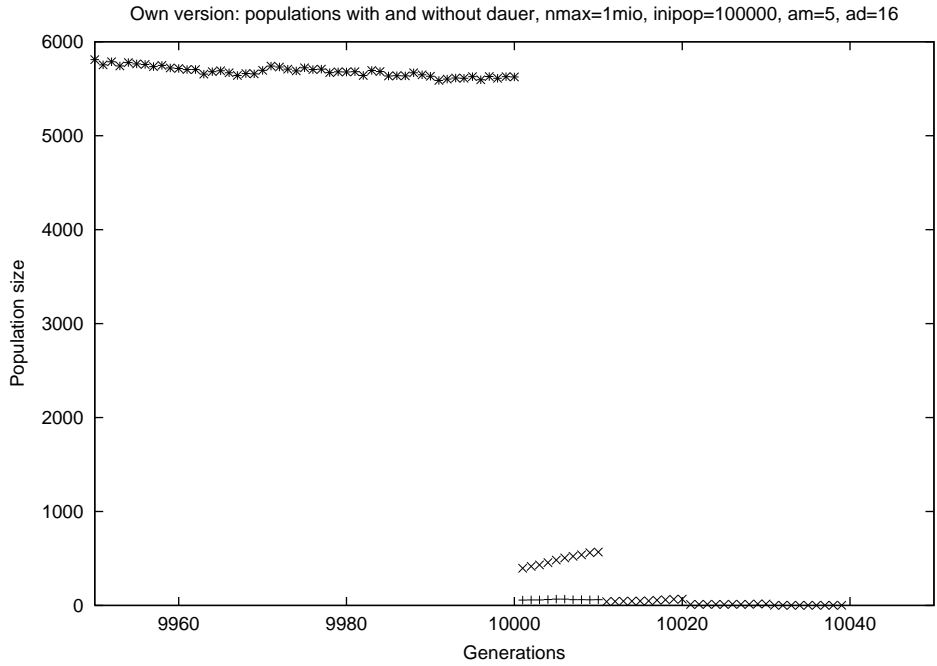}
\caption{After 10000 generations, in each tenth summer only 1 \% survive.
'x'=population with dauer, $s_b=0.01$, '+'=population without dauer, $s_b=0.01$}
\end{figure}

\begin{figure}[h]
\centering
\includegraphics[width=9cm, angle=-90]{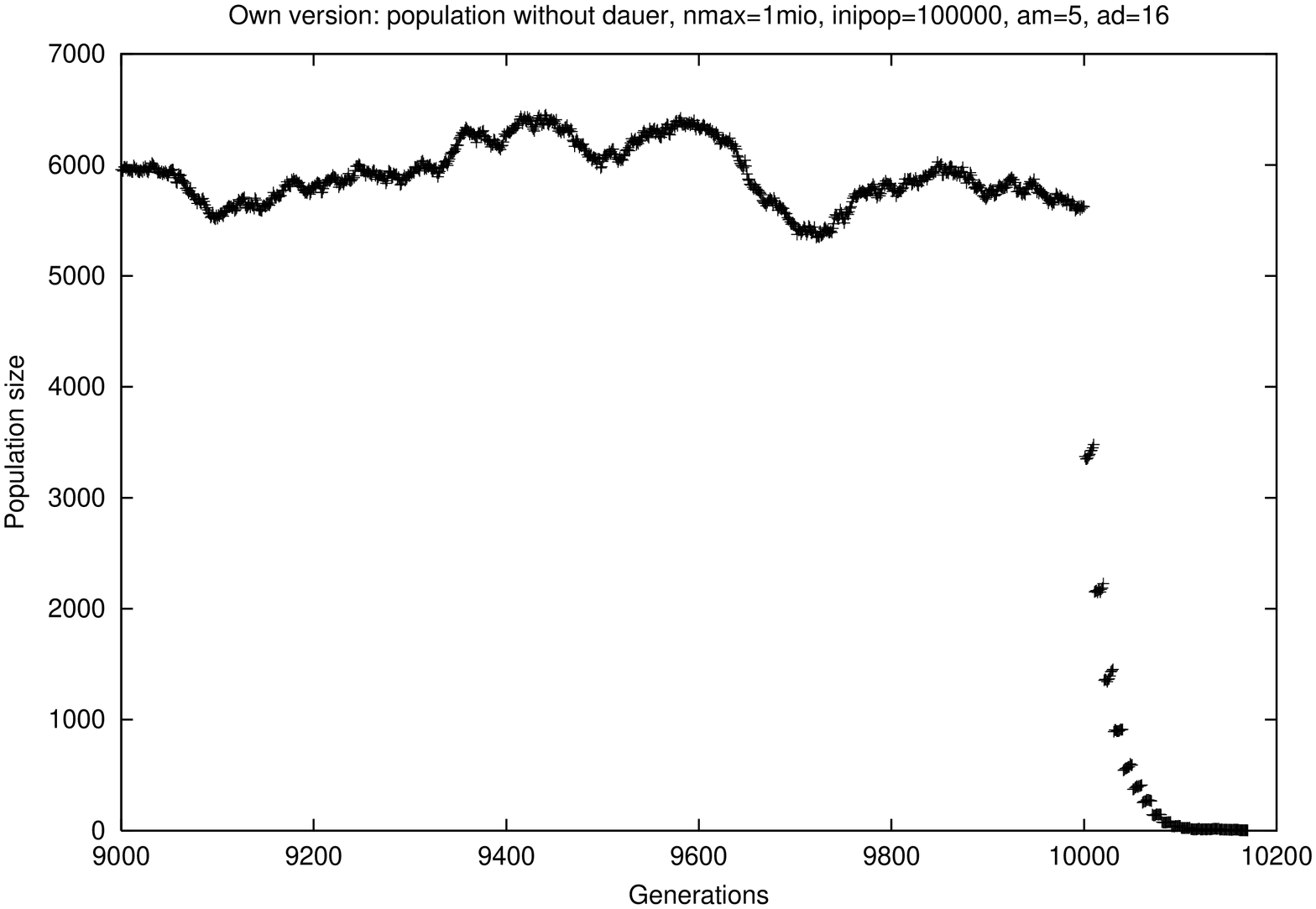}
\caption{After 10000 generations, in each tenth summer 61 \% survive.}
\end{figure}

\clearpage
\begin{figure}[!h]
\centering
\includegraphics[width=9cm, angle=-90]{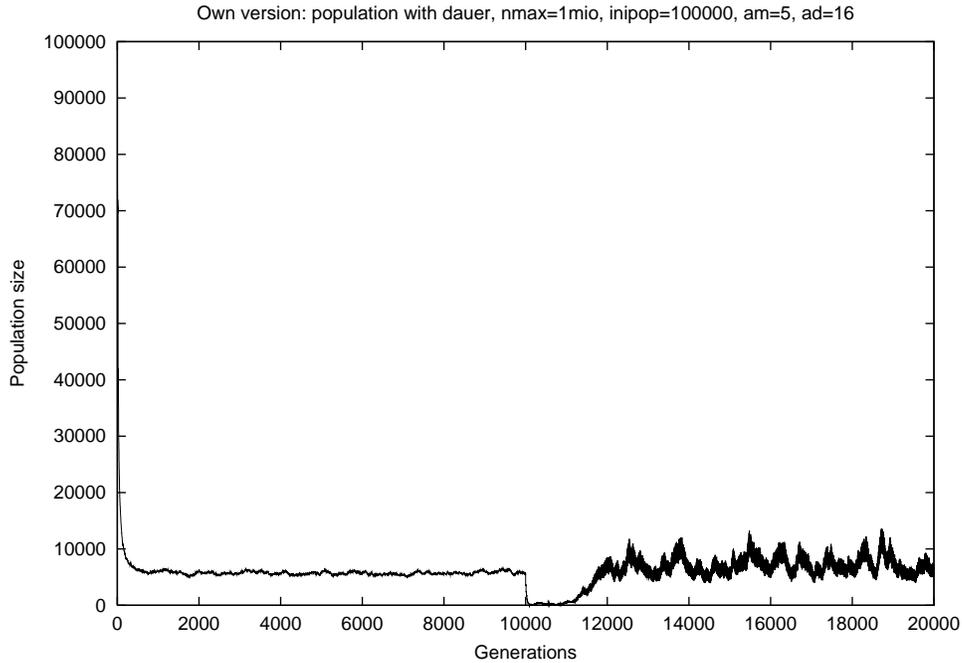}
\caption{After 10000 generations, in each tenth summer 61 \% survive.}
\end{figure}

\subsection{The Stauffer model with increased birth rate}
This difference in the two versions introduces the question which parameters lead to a preservation of the \emph{dauer population} in Stauffers interpretation. As increasing the \emph{bad summer factor} $s_b$ up to $s_b=0.9$ still kills the population with \emph{dauer mutation}, a change in the birth rate might show the advantage of the \emph{dauer larvae} for the preservation of the species. Iterating the birth loop twice in Stauffers version, provides the results in figure 7 for $s_b=0.23$. Whereas the population with \emph{dauer mutation} survives the bad conditions, the population without it dies out (the small dots ending near generation 10000).
\\Analysis showed that for $s_b=0.23$ and higher the population with \emph{dauer larvae} survives, whereas the population without it dies out. Is $s_b<0.23$ both populations are extincted. Thus an increased birth rate in Stauffers version leads to the preservation of the species for $s_b=0.23$ and higher.

\begin{figure}[!h]
\centering
\includegraphics[width=9cm, angle=-90]{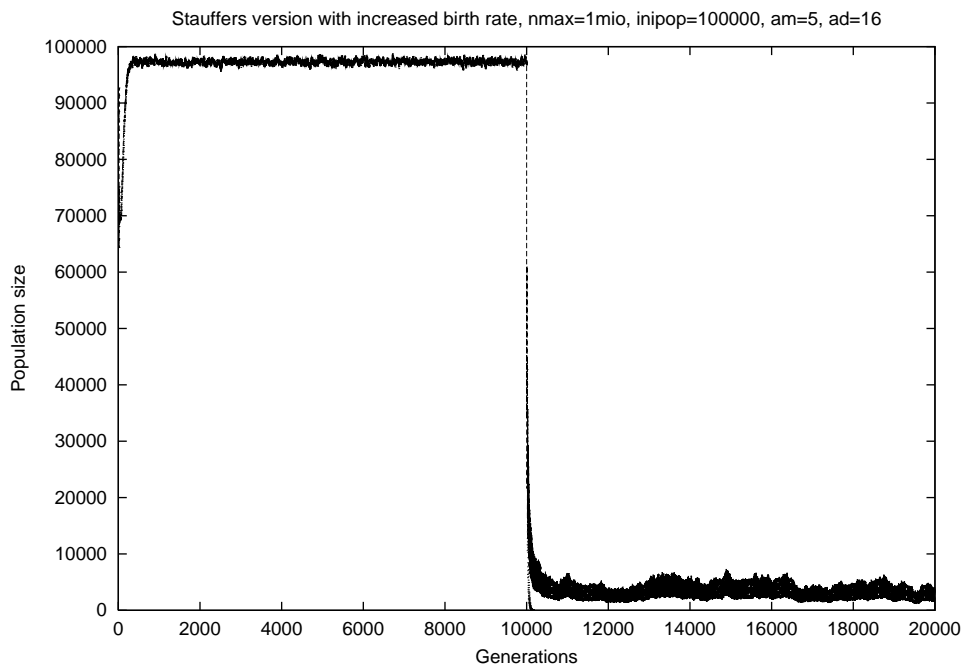}
\caption{After 10000 generations, in each tenth summer 23 \% survive.
'lines'=population with dauer, $s_b=0.23$, 'dots'=population without dauer, $s_b=0.23$}
\end{figure}

\end{document}